\documentclass[preprint,12pt]{elsarticle}

\journal{Nuclear Physics A}

\usepackage{multirow}
\usepackage{caption}
\usepackage{subcaption}
\usepackage{graphicx}
\usepackage{longtable}
\usepackage{amssymb}
\usepackage{enumerate}

\begin{document}
\newcommand{\be}{{B(E2) }}
\newcommand{\ave}[1]{\ensuremath{\left\langle #1 \right\rangle}}
\newcommand{\var}[1]{\ensuremath{\mathrm{Var}\left( #1 \right)}}
\newcommand{\ggp}{NRF}
\newcommand{\ee}{\ensuremath{e, e^\prime} }
\newcommand{\nuc}[2]{\ensuremath{^{#1}}{#2}}

\begin{frontmatter}
\title{On the Equivalence of  Experimental B(E2) Values Determined by Various Techniques}
\author[label1]{M. Birch}
\author[label2]{B. Pritychenko\corref{cor1}}
\ead{pritychenko@bnl.gov}
\author[label1]{B. Singh}
\address[label1]{Department of Physics \& Astronomy, McMaster University, Hamilton, \\ Ontario L8S 4M1, Canada}
\address[label2]{National Nuclear Data Center, Brookhaven National Laboratory, \\ Upton, NY 11973-5000, USA}
\cortext[cor1]{Corresponding author}
\begin{abstract}
We establish the equivalence of the various techniques for measuring \be values using a statistical analysis.
Data used in this work come from the recent compilation by B. Pritychenko {\it et al.},  {\sc At. Data Nucl. Data Tables} 107 (2016).
We consider only those nuclei for which the \be values were measured by at least two different methods,
with each method being independently performed at least twice. Our results indicate that most prevalent methods of measuring
\be values are equivalent, with some weak evidence that  Doppler-shift attenuation method (DSAM)
measurements may differ from Coulomb excitation (CE) and nuclear resonance fluorescence (NRF) measurements. However,
such an evidence appears to arise from discrepant DSAM measurements of the lifetimes for \nuc{60}{Ni} and some Sn nuclei rather than
a systematic deviation in the method itself.
\end{abstract}


\end{frontmatter}

\section{Introduction}
Reduced transition probabilities, the \be values for transitions from the first excited $2^{+}$ to the ground states in even-even nuclei are
fundamentally important quantities in nuclear physics for determining the collectivity in nuclei. As such, experimental efforts to
measure \be values have been going on for the past sixty years and many different techniques have been developed
for this purpose. The two most common methods are Coulomb excitation (CE) and Doppler-shift
attenuation (DSAM). A recent CE study of the \be values for Sn nuclei \cite{2015Al24} compared their
measurements with previous works and revealed an apparent systematic disagreement with a set of DSAM measurements
\cite{2011Ju01}. Without doubting the standard formulations relating \be and lifetime, this does raise the question
that, given the different sets of systematic errors associated with different techniques, are the results
significantly impacted
such that the methods may or may not be considered equivalent.
To the best of our knowledge, this question has never been
addressed comprehensively across the entire chart of nuclides while comparing different methods, although, limited comparisons have been
published between CE experiments at different energies \cite{2006Co03,2008Sc22}.

Our newly published compilation and evaluation of experimental \be values \cite{2016Pr01} for the first
$2^{+}$ states in the even-even nuclei, an extensive update of the
previous work by Raman et al. \cite{2001Ra27}, provides an opportunity to adequately answer
the question of whether systematic differences exist between different methods of determining \be values.
In particular, we focus here on the most common experimental techniques for measuring \be values, namely:
Doppler-shift attenuation method (DSAM), recoil distance Doppler-shift (RDDS),
delayed coincidences (DC), Coulomb excitation (CE), nuclear resonance fluorescence (\ggp) and the weakly model
dependent method of electron scattering (\ee). Note that DSAM includes also ``transmission'' Doppler-shift attenuation
in which particles are still travelling through the medium when gamma-rays are detected.
In making comparisons between methods, using our
compiled data, we have selected only those nuclei which have had their \be values measured by at least two
different methods, with each method being independently employed by different laboratories at least twice. This requirement implies that
we can always take an average when determining the result from a particular method, reducing the likelihood
of an outlying measurement having a large influence on our conclusions. As a result, of the 447 nuclei which
were included in our compilation, we can only use measurements on 100 of those nuclei. Table~\ref{tab:be2data}
lists these nuclei and their \be values as determined by the different experimental methods.

\section{Analysis}
Our statistical analysis is performed as follows:
let $x \pm \delta x$ and $y \pm \delta y$ be two measurements of the \be value of the
same nucleus determined by different methods. Note that since we insist that each experimental
method be performed at least twice on each isotope, $x$ and $y$ are each an average
of at least two independent measurements. We define the normalized difference between
the two results, $x \pm \delta x$ and $y \pm \delta y$, as
\begin{equation}\label{eq:normdiff}
z = \frac{x - y}{\sqrt{\delta x^2 + \delta y^2}}.
\end{equation}
Assume that $x$ was sampled from a random variable, $X$, which is normally
distributed such that \ave{X} is the ``true value'' of the \be value for the
nucleus and $\var{X} = \delta x^2$. Similarly, assume $y$ was sampled from
a normally distributed random variable, $Y$, such that $\var{Y} = \delta y^2$.
The two experimental procedures will be equivalent for this nucleus if
$\ave{X} = \ave{Y}$. Under these assumptions, $z$ will be sampled from
a normal distribution with mean given by
\begin{equation}
\ave{Z} = \frac{\ave{X} - \ave{Y}}{\sqrt{\delta x^2 + \delta y^2}}
\end{equation}
and unit variance. Clearly, if the two experimental methods are
equivalent, then $\ave{Z} = 0$, independent of which nucleus is being considered. Also notice that
since $z$ is dimensionless and $\var{Z} = 1$, we can interpret $z$ as being
the difference between $x$ and $y$ in units of standard deviations of $(X - Y)$.
Now consider two sets of \be measurements, $\{ x_i \pm \delta x_i \}_{i=1}^N$ and
$\{ y_i \pm \delta y_i \}_{i=1}^N$. Each $x_i$ is determined by one experimental
method, while each $y_i$ is determined by another. The two sets are paired such
that $x_i$ and $y_i$ are measurements on the same nucleus and hence for each
such pair we can calculate $z_i$ according to equation~(\ref{eq:normdiff}).
Under the hypothesis that the two methods are equivalent, $\{z_i\}_{i=1}^N$
is a sample of size $N$ from a standard normal distribution (i.e. one with
zero mean and unit variance). However, as we shall show later, the actual
distribution of $\{z_i\}_{i=1}^N$ we observe has some degree of asymmetry.
This could arise due to underestimated uncertainties since that would artificially
increase the resulting values of $z$. We circumvent this problem phenomenologically
by postulating (and later provide evidence using the real data) that the distribution of
$\{z_i\}_{i=1}^N$ is in fact an asymmetric normal distribution, with probability density function given by
\begin{equation}\label{eq:asymnorm}
f(z;m,a,b) = \left\{ \begin{array}{c}
\sqrt{\frac{2}{\pi(a + b)^2}} \exp\left( -\frac{1}{2} \frac{(z - m)^2}{b^2} \right), ~z \le m \\
\sqrt{\frac{2}{\pi(a + b)^2}} \exp\left( -\frac{1}{2} \frac{(z - m)^2}{a^2} \right), ~z > m
\end{array} \right.,
\end{equation}
where $m$ is the mode (most probable value) of the distribution, and $a$, $b$ give the upper and lower standard
deviations, respectively. By fitting the values of $m$, $a$ and $b$ to the distribution of
$\{z_i\}_{i=1}^N$, we can compute the
probability that methods $x$ and $y$ will differ by $n$ standard deviations
as (recall the interpretation of $z$ above)
\begin{equation}\label{eq:pn}
P_n = \Pr( \vert z \vert > n ) = \int_{-\infty}^{-n} f(z;m, a, b) dz + \int_n^\infty f(z;m, a, b) dz,
\end{equation}
the average difference as
\begin{equation}\label{eq:mud}
\mu_d = \int_{-\infty}^\infty z f(z;m, a, b)dz = m + \sqrt{\frac{2}{\pi}} (a-b),
\end{equation}
and the most probable deviation is simply $m$. Each of these quantities provides a different measure of difference
between methods $x$ and $y$. Using these measures we define three different criteria for determining that two
methods are not equivalent:
\begin{enumerate}[(i)]
\item $P_1 > 0.5$, i.e. it is more probable that $(x-y)$ differs from zero by at least one standard deviation than not;
\item $\vert\mu_d\vert > 1$, i.e. on average $(x-y)$ differs from zero by at least one standard deviation;
\item $\vert m\vert > 1$, i.e. the most probable value of $(x-y)$ differs from zero by more than one standard deviation.
\end{enumerate}
Criteria (ii) and (iii) also give information regarding the direction of the deviation, e.g. if $\mu_d < -1$ then
not only can we say that $x$ is not equivalent to $y$, but that it is on average less than $y$.

\section{Results}
\begin{figure}
\centering
\begin{subfigure}{0.75\textwidth}
  \centering
  \includegraphics[width=\linewidth]{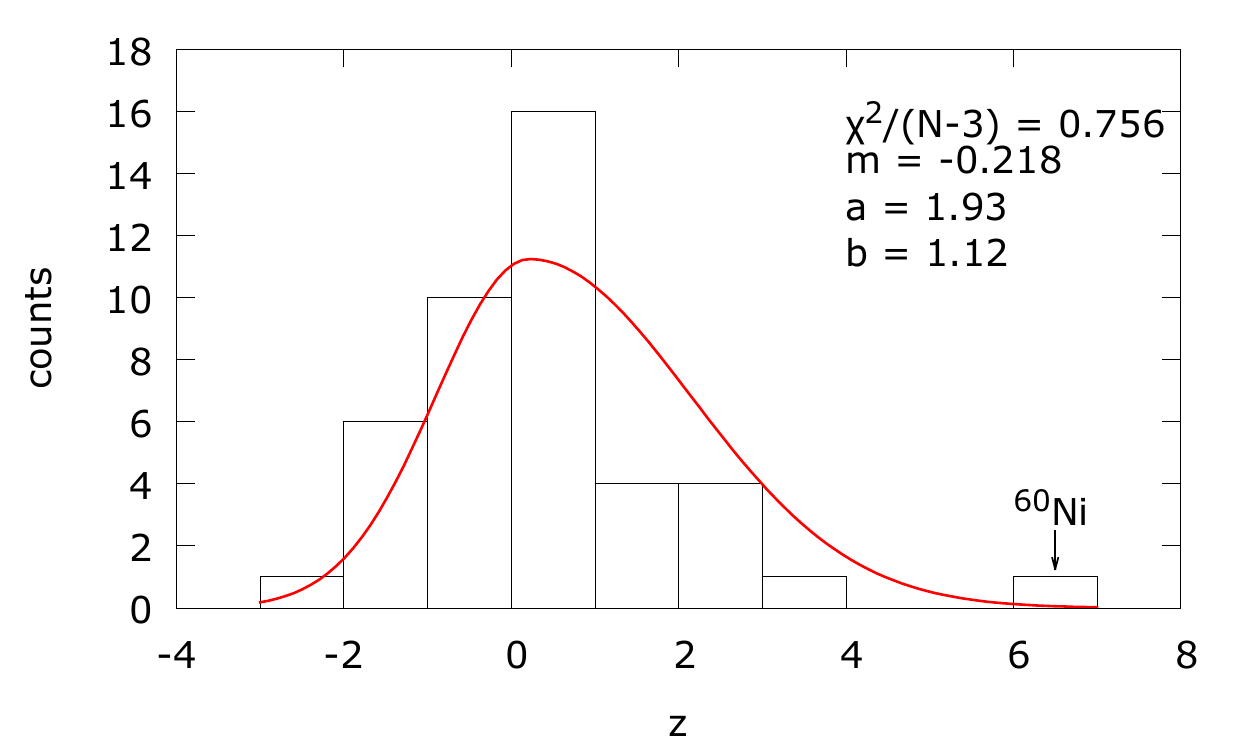}
  \caption{\label{fig:cedsa}}
\end{subfigure}
\begin{subfigure}{0.75\textwidth}
  \centering
  \includegraphics[width=\linewidth]{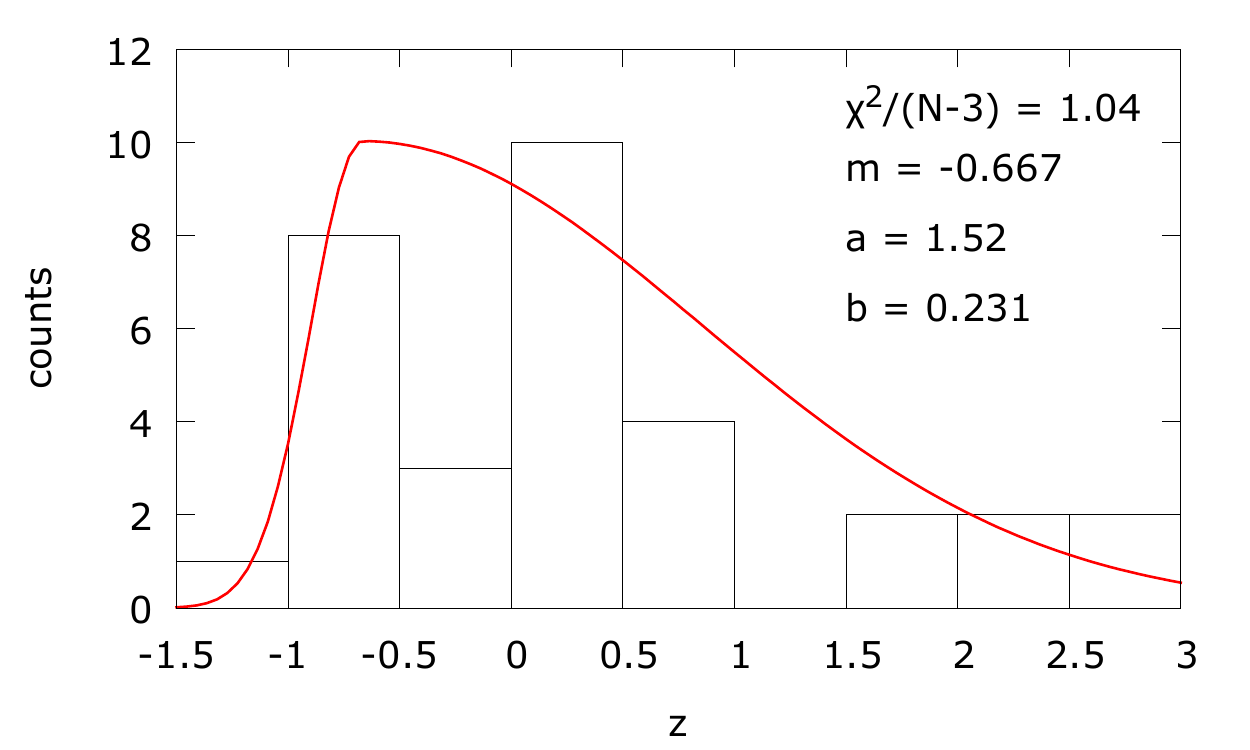}
  \caption{\label{fig:ceee}}
\end{subfigure}
\caption{(a) Histograms of the $z$ values computed for the CE/DSAM method pair, together with the asymmetric
normal distribution fit. The \nuc{60}{Ni} outlier in this dataset is also indicated. (b) Same as (a), but
for the CE/(\ee) pair.\label{fig:hists}}
\end{figure}

Let us consider a particular example to illustrate this analysis in detail. The two most common methods for
determining \be values are CE and DSAM, so let us compare those methods (i.e. choose $x=$ CE and $y=$ DSAM).
There are 43 nuclei which have at least two independent DSAM and two
independent CE measurements. We can compute the normalized differences between the two methods, $z_i$, as
defined above for these 43 cases. To show that these data are not normally distributed, we performed a Shapiro-Wilk test
\cite{1995RoAA}. This test uses a statistic denoted $W$ which is close to unity for normally distributed
data and close to zero otherwise. For this dataset, we found that $W = 0.928$, which gives a
p-value of $0.0096$, i.e. the probability of obtaining that value of $W$,
assuming that the data are normally distributed, is less than 1\%. Therefore, it is statistically
significant that $\{ z_i \}_{i=1}^{43}$ are not normally distributed. Instead we can fit an asymmetric normal
distribution to $\{ z_i \}_{i=1}^{43}$ and obtain $m=-0.218$, $a=1.93$, $b=1.12$. Using Pearson's $\chi^2$ test for
goodness of fit, $\chi^2/(N - 3) = 0.756$, which is smaller than the critical $\chi^2$ for rejection at
99\% confidence level, indicating an acceptable fit to the data. Hence we have shown, as claimed
above, that $z_i$ is not normally distributed, but is well-modelled as being distributed according to an
asymmetric normal distribution. A histogram of $\{z_i\}_{i=1}^{43}$ together with the asymmetric normal
fit is shown in Fig.~\ref{fig:cedsa}. Using this fit we can calculate the various quantities defined above to test for
equivalence between CE and DSAM: $P_1 = 0.513$, $\mu_d = -0.428$ and  $m=-0.218$. Therefore,
we would conclude that DSAM and CE are equivalent methods using criteria (ii) and (iii), however
they are not equivalent by criterion (i). This discrepancy is further discussed in Section \ref{sec:dis}.
Fig.~\ref{fig:cedsa} could be misleading in the sense that the remaining distribution, after
removing the outlying point, appears symmetric and so one might concluded that using an
asymmetric distribution in this analysis is not actually necessary. To clarify this point, Fig.~\ref{fig:ceee}
shows a histogram of the $z$ values for the CE/(\ee) pair. In this case $P_1 = 0.256$,
$\mu_d = 0.362$ and  $m=-0.667$, hence the two methods are equivalent according to all three
criteria. However, we see that the distribution is still quite asymmetric. The histograms of
$z$ values for CE with the remaining methods (DC, RDDS, \ggp) are given in \ref{app:hist}
also show asymmetric distributions.

\begin{figure}
\centering
\begin{subfigure}[t]{0.4\textwidth}
  \centering
  \includegraphics[width=\linewidth]{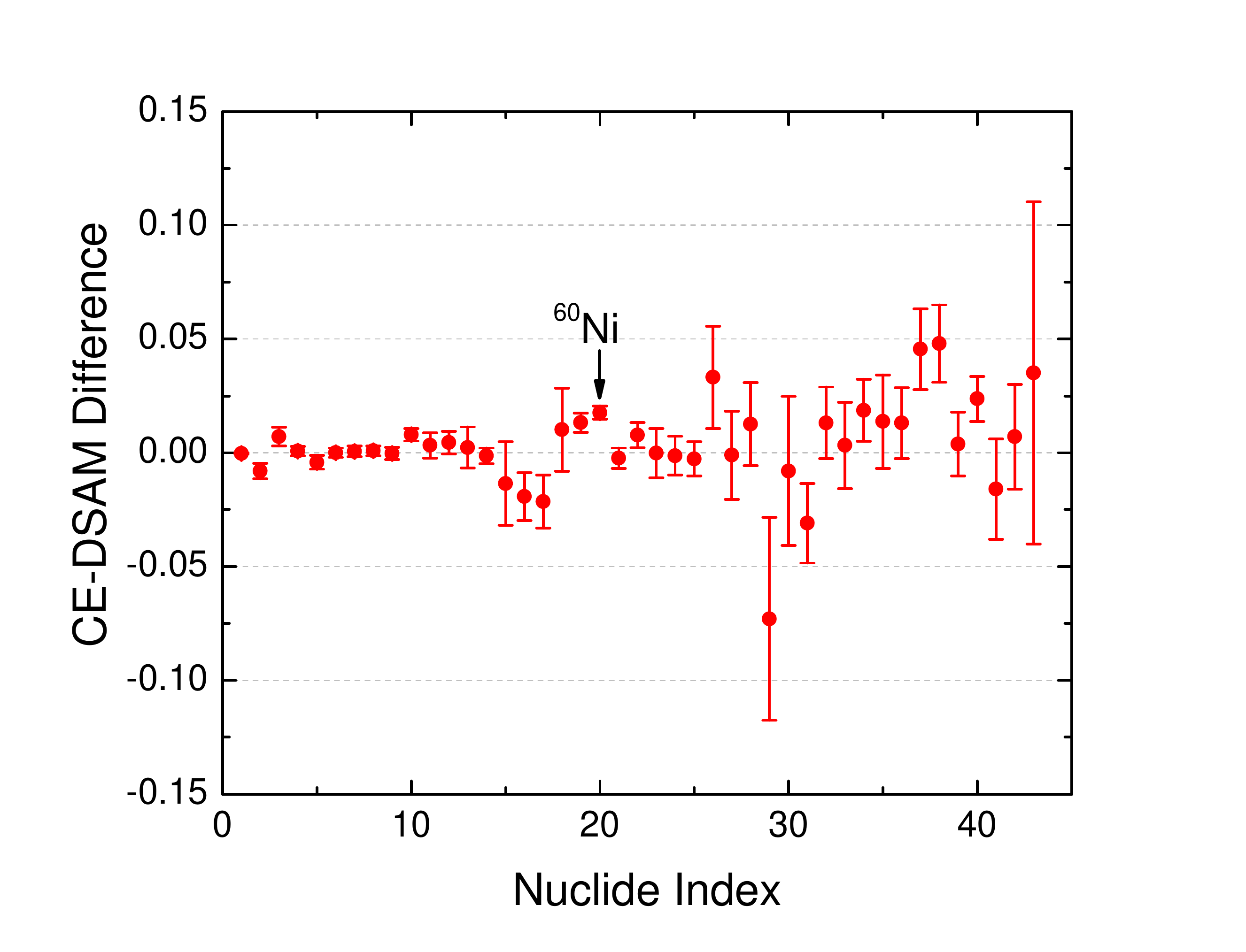}
  \caption{\label{fig:diffs_dsa}}
\end{subfigure}
~
\begin{subfigure}[t]{0.4\textwidth}
  \centering
  \includegraphics[width=\linewidth]{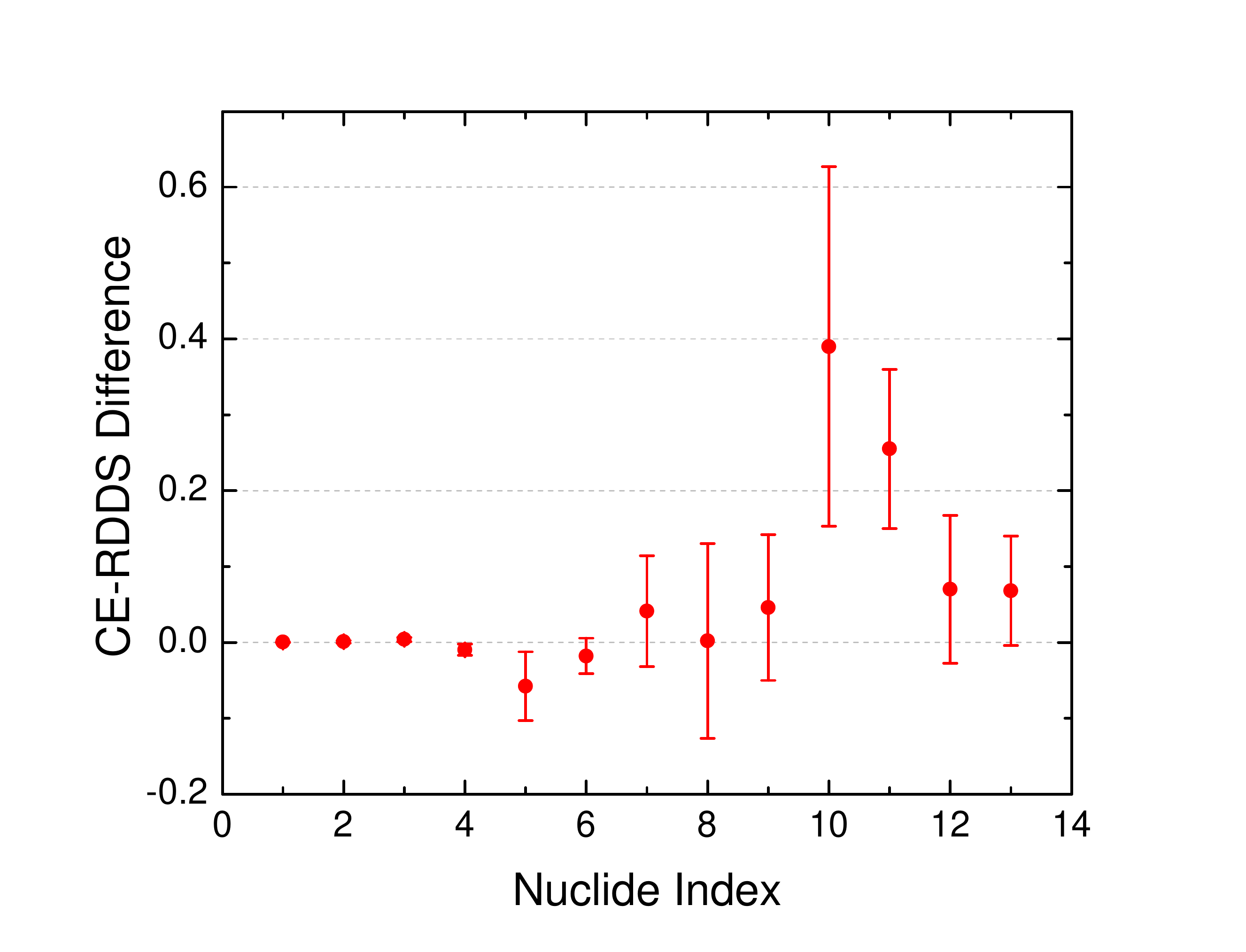}
  \caption{\label{fig:diffs_rdds}}
\end{subfigure}
~
\begin{subfigure}[t]{0.4\textwidth}
  \centering
  \includegraphics[width=\linewidth]{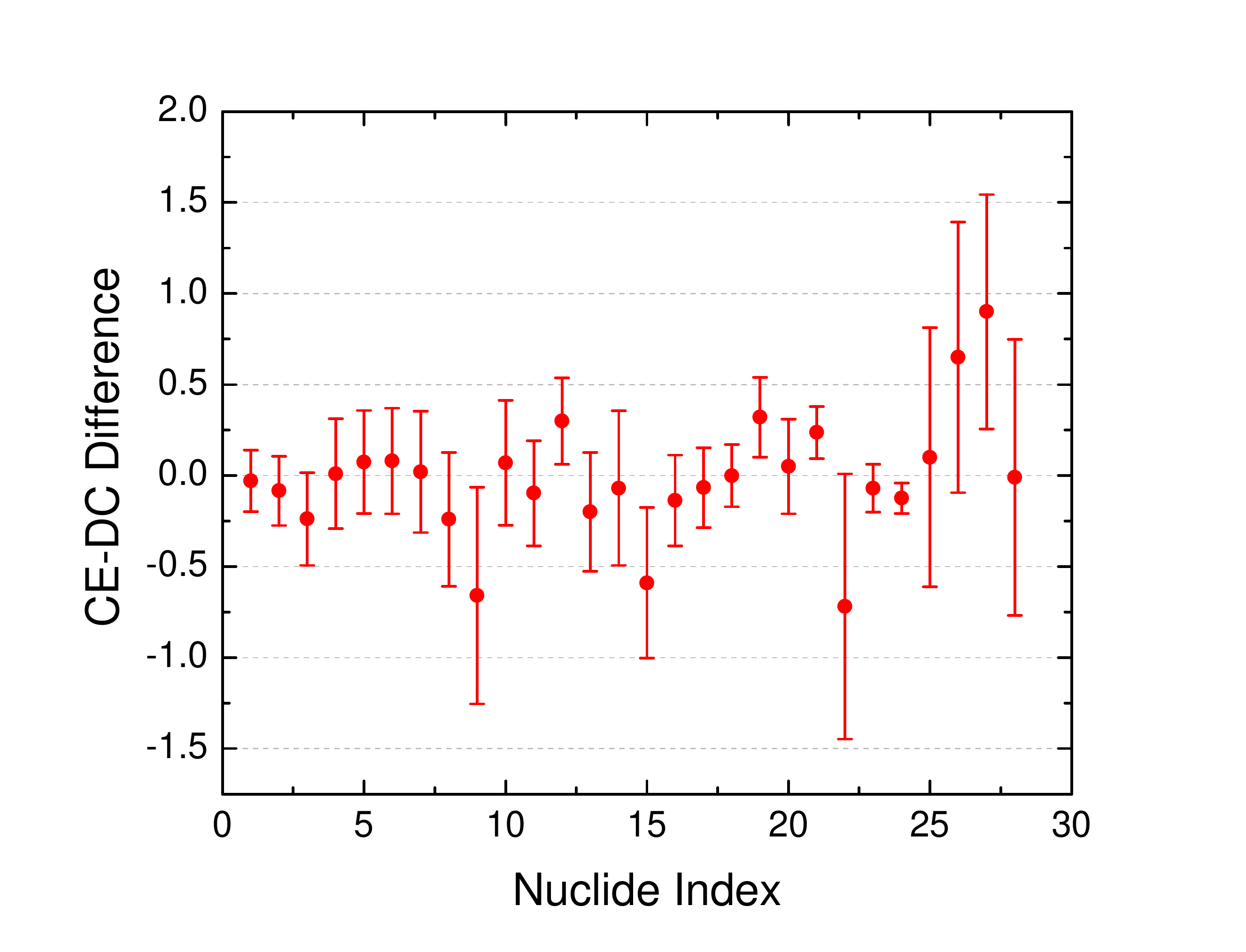}
  \caption{\label{fig:diffs_dc}}
\end{subfigure}
~
\begin{subfigure}[t]{0.4\textwidth}
  \centering
  \includegraphics[width=\linewidth]{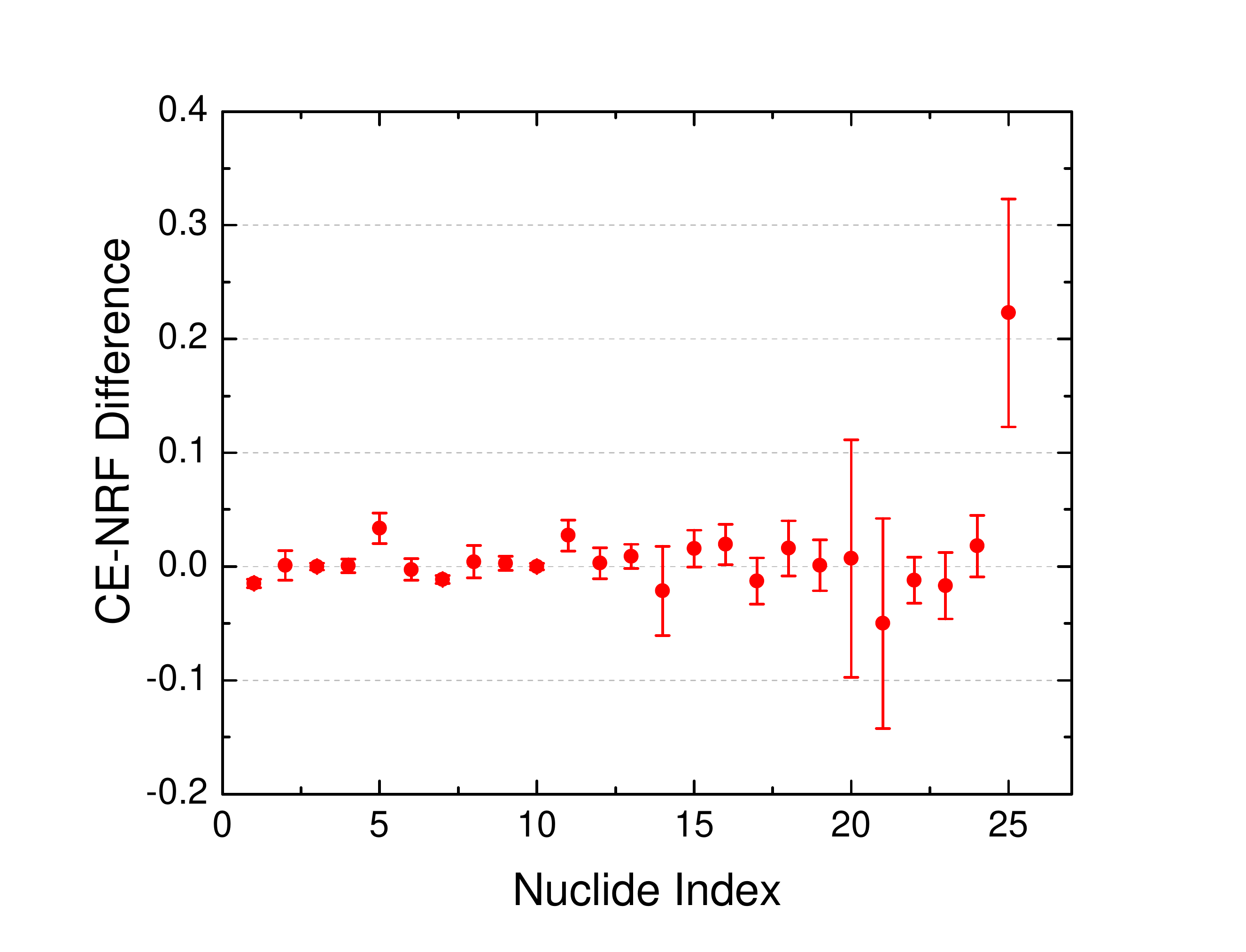}
  \caption{\label{fig:diffs_gg}}
\end{subfigure}
~
\begin{subfigure}[t]{0.4\textwidth}
  \centering
  \includegraphics[width=\linewidth]{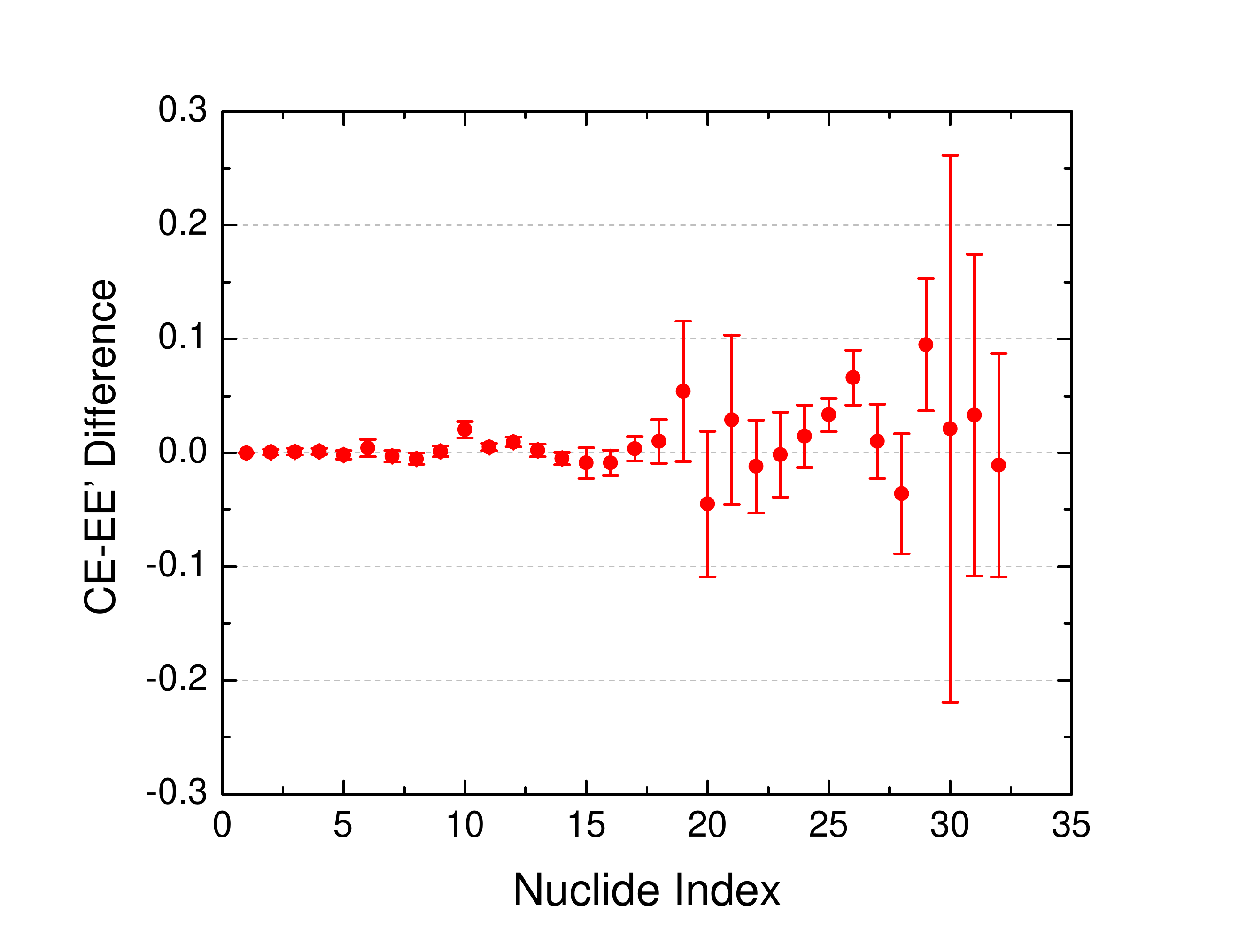}
  \caption{\label{fig:diffs_ee}}
\end{subfigure}
\caption{Difference between \be values determined by CE and each other method: (a), (b), (c), (d) (e)
compare with DSAM, RDDS, DC, NRF and (\ee) respectively. The x-axis in each plot
shows the index associated with the nuclide for that pairing of methods. These indices are essentially
arbitrary and differ between the different plots, but increasing index always corresponds to increasing
mass number. See Table~\ref{tab:indicies} to see which nuclide corresponds to each index.
\label{fig:diffs}}
\end{figure}

In order to systematically establish the equivalence of each method to one another, we performed this
analysis for each pair of methods and did not find any pairs which satisfied criteria (ii) or (iii), although
a DSAM/CE and DSAM/NRF did satisfy (i). However, in both of these cases
neither of the other two criteria are also satisfied. This gives some evidence that there may be
problems with some of DSAM measurements, but not necessarily anything systematically wrong with the
method itself. Indeed, if the DSAM measurements for \nuc{60}{Ni} are excluded then the discrepancy with
CE and NRF no longer exists. This is discussed in detail in Section \ref{sec:dis}. In this global comparison,
we also do not have any evidence for a systematic deviation of \be values in Sn nuclides as suggested in the
recent study \cite{2015Al24}.

Since CE is the most commonly used method, and presumably it suffers less from systematic errors such as
stopping powers in DSAM method, detailed results for comparing CE with several other methods
are shown in Table~\ref{tab:results}. Table~\ref{tab:results} also further illustrates the asymmetry of the
$z$-value distributions. For a symmetric distribution (i.e. usual normal distribution) $\mu_d = m$,
yet in the table we see the mean and mode differ significantly in all cases.
The comparison between CE and each of the other methods can also be
seen graphically in Fig.~\ref{fig:diffs} where the difference (not normalized to the
size of the error bars, i.e. $(x-y)$ and not $z$) between the \be
values measured by CE and each of the other methods are plotted.

\begin{table}
\caption{Comparison between CE and five other methods using each of the metrics outlined in equations
(\ref{eq:asymnorm}), (\ref{eq:pn}) and (\ref{eq:mud}) and used to define non-equivalence
criteria (i)-(iii). The quantity $\chi^2/(N-3)$ gives the reduced $\chi^2$
values for the fit of an asymmetric normal distribution to the calculated $\{ z_i \}_{i=1}^N$,
establishing the goodness of fit.\label{tab:results}}
\begin{center}
\begin{tabular}{cccccc} \hline\hline
Method & $\chi^2/(N-3)$ & $P_1$ & $\mu_d$ & $m$ \\ \hline
DSAM$^\ast$ & 0.756 & 0.513 & 0.428 & $-0.218$ \\
RDDS & 0.491 & 0.423 & 0.484 & 0.697 \\
DC & 0.533 & 0.229 & 0.0156 & $-0.220$ \\
NRF & 2.51 & 0.444 & $-0.236$ & 0.116 \\
\ee & 1.04 & 0.256 & 0.362 & $-0.667$ \\
\hline\hline
\end{tabular}
\end{center}
$^\ast$ {\tiny If the \nuc{60}{Ni} DSAM measurements are excluded then $P_1 = 0.469$, $\mu_d = 0.264$, $m = 0.0944$.}
\end{table}

\section{Discussion and Conclusion}\label{sec:dis}
The results of Table~\ref{tab:results} overall indicate that each method used to measure \be values is
equivalent to CE, however some points are worth further discussion. All the $\chi^2/(N-3)$ values are smaller
than the critical $\chi^2$ values for rejection at 99\% confidence except NRF, i.e. only the CE/NRF
normalized differences are not well-modelled by an asymmetric normal distribution. However, a visual
inspection of Fig.~\ref{fig:diffs_gg} reveals that, even if the numeric values of the parameters for
NRF in Table~\ref{tab:results} are not meaningful, the conclusion of equivalence is still supported.
Indeed, about 75\% of the CE/NRF differences lie within one standard deviation of zero.

The only indication of non-equivalence in Table~\ref{tab:results} is $P_1 > 0.5$ for CE/DSAM. Although not listed,
it is also true that $P_1 > 0.5$ for NRF/DSAM as well. This could indicated a systematic problem with DSAM
measurements, however the reason for these discrepancies can be attributed to measurements on a single
nucleus, \nuc{60}{Ni}. As can be seen in the histogram of Fig.~\ref{fig:cedsa} a single $z$ value is separated
from the bulk of the distribution by at least a full unit (standard deviation). This data point is also indicated
in Fig.~\ref{fig:diffs_dsa} and one can see that indeed the error bars are very small compared to its deviation
from zero (hence the large value of $z$ for this measurement). This disagreement between CE and DSAM measurements
was also noted in \cite{2014Al20}, although very good agreement was found for \nuc{62}{Ni}.
Further evidence against the DSAM
lifetime measurements of \nuc{60}{Ni} is the fact that they are in mutual disagreement with one another.
The more recent measurements give $\sim$1.3~ps
\cite{2001Ke02,2008Or02} and earlier measurements $\sim$1~ps \cite{1973Fi15,1973Ro20}.
This could be due to problems with one or more of the DSAM measurements of this nucleus
as a result of poorly determined stopping powers, improperly implemented inverse kinematics
during data analysis or ignored possible angular correlations. In any case, it appears that
the \nuc{60}{Ni} DSAM measurements are an isolated problem rather than evidence of a broad systematic issue
with the method. By excluding these measurements from the analysis, there is no longer significant evidence that
DSAM disagrees with CE or NRF.

In conclusion, based on the analyses performed on the most recently compiled \be experimental data \cite{2016Pr01},
we conclude that the most commonly methods used in the measurement of \be values are equivalent. There is some
evidence to suggest that
there is disagreement between DSAM and CE as well as DSAM and NRF methods, however these appear to arise from
discrepant DSAM measurements in the lifetime of
\nuc{60}{Ni} and not from a systematic deficiency in the method itself. Similarly, we do not find evidence for
systematic deviation between DSAM and CE results for Sn nuclei, when all the available experimental data are
considered. It is possible that DSAM results reported in \cite{2011Ju01} suffer from some experimental difficulties.

\section{Acknowledgments}
\label{sec:Acknowledgements}
We are indebted to Dr. M. Herman (BNL) for support of this project. The authors gratefully
acknowledge J.M. Allmond (ORNL) for providing his latest results and fruitful discussions.
Work at Brookhaven was funded by the Office of Nuclear Physics, Office of Science of the U.S. Department
of Energy, under Contract No. DE-AC02-98CH10886 with Brookhaven Science Associates, LC. Work at
McMaster University was partially supported by the Office of Science of the U.S. Department
of Energy.

\newpage
\appendix
\setcounter{table}{0}
\renewcommand{\thetable}{A\arabic{table}}
\section{Table of Data}

\begin{center}
\begin{longtable}{c|p{1.9cm}p{1.9cm}p{1.9cm}p{1.9cm}p{1.9cm}p{1.9cm}}
\caption{\be recalculated weighted averages for DSAM, RDDS, DC, CE, NRF and \ee experimental  methods using the relevant compiled data in 
Table 1 of 2016 compilation \cite{2016Pr01}.\label{tab:be2data}}
%\noalign \\
\\
\hline \hline
 \multirow{2}{*}{\textbf{Nuclei}}  & \multicolumn{6}{c}{\textbf{B(E2)$\uparrow$ ($e^{2}b^{2}$)}} \\
                  \cline{2-7}
   &  \textbf{DSAM} &  \textbf{RDDS} &  \textbf{DC} &  \textbf{CE} &  \textbf{NRF} &  \textbf{\ee} \\
\hline \\
\endfirsthead

\hline \hline
 \multirow{2}{*}{\textbf{Nuclei}}  & \multicolumn{6}{c}{\textbf{B(E2)$\uparrow$ ($e^{2}b^{2}$)}} \\
                  \cline{2-7}
   &  \textbf{DSAM} &  \textbf{RDDS} &  \textbf{DC} &  \textbf{CE} &  \textbf{NRF} &  \textbf{\ee} \\
\hline  \\
\endhead

\nuc{12}{C} & 0.00400(20) &  &  &  & 0.00334(71) & 0.00399(33) \\
\nuc{16}{O} &  &  &  &  & 0.00371(32) & 0.00413(36) \\
\nuc{18}{O} & 0.00458(24) & 0.00405(20) &  & 0.00429(15) &  & 0.00449(22) \\
\nuc{18}{Ne} & 0.0241(22) &  &  & 0.0160(26) &  &  \\
\nuc{20}{Ne} & 0.0299(28) &  &  & 0.0369(30) &  &  \\
\nuc{22}{Ne} & 0.0230(15) & 0.0229(11) &  & 0.0237(14) &  & 0.0232(22) \\
\nuc{24}{Mg} & 0.0472(23) & 0.0391(19) &  & 0.0430(20) & 0.0578(32) & 0.0421(23) \\
\nuc{26}{Mg} & 0.0309(15) &  &  & 0.0309(15) & 0.030(13) & 0.0298(21) \\
\nuc{28}{Si} & 0.0323(16) &  &  & 0.0329(17) & 0.0330(27) & 0.0348(31) \\
\nuc{32}{S} & 0.0293(14) &  &  & 0.0301(16) & 0.0296(58) & 0.0259(74) \\
\nuc{34}{S} & 0.0210(12) &  &  & 0.0207(24) &  &  \\
\nuc{36}{Ar} & 0.0219(16) &  &  & 0.0298(23) &  &  \\
\nuc{40}{Ar} & 0.0327(24) &  &  & 0.0359(50) &  &  \\
\nuc{40}{Ca} & 0.0093(10) &  &  &  & 0.0103(11) & 0.00703(82) \\
\nuc{42}{Ca} & 0.0334(24) &  &  &  & 0.060(12) & 0.0367(49) \\
\nuc{44}{Ca} & 0.0440(36) &  &  & 0.0485(35) &  & 0.0517(35) \\
\nuc{48}{Ca} & 0.0085(13) &  &  &  &  & 0.00826(50) \\
\nuc{46}{Ti} &  & 0.0995(48) &  & 0.0898(60) & 0.056(12) &  \\
\nuc{48}{Ti} & 0.0626(55) &  &  & 0.0649(72) & 0.0676(60) &  \\
\nuc{50}{Ti} & 0.0281(19) &  &  & 0.0267(29) &  &  \\
\nuc{50}{Cr} & 0.118(18) &  &  & 0.1045(32) &  &  \\
\nuc{52}{Cr} & 0.077(10) &  &  & 0.0573(31) & 0.0687(13) & 0.0627(39) \\
\nuc{54}{Fe} & 0.077(11) &  &  & 0.0551(38) &  & 0.0539(24) \\
\nuc{56}{Fe} & 0.088(18) &  &  & 0.0981(26) & 0.094(14) & 0.0777(67) \\
\nuc{58}{Ni} & 0.0560(37) &  &  & 0.0692(20) & 0.0665(59) & 0.0643(25) \\
\nuc{60}{Ni} & 0.0750(21) &  &  & 0.0926(20) & 0.0927(20) & 0.0832(37) \\
\nuc{62}{Ni} & 0.0908(34) &  &  & 0.0884(30) &  & 0.0862(46) \\
\nuc{64}{Ni} & 0.0590(40) &  &  & 0.0668(38) &  & 0.0720(41) \\
\nuc{64}{Zn} & 0.1512(42) &  &  & 0.151(10) & 0.1237(93) & 0.1601(90) \\
\nuc{66}{Zn} & 0.1379(29) &  &  & 0.1366(80) & 0.134(11) & 0.1454(80) \\
\nuc{68}{Zn} & 0.1206(26) &  &  & 0.1179(70) & 0.1091(81) & 0.1145(80) \\
\nuc{70}{Zn} & 0.1429(80) &  &  & 0.176(21) &  &  \\
\nuc{70}{Ge} & 0.179(19) &  &  & 0.1779(35) &  &  \\
\nuc{72}{Ge} & 0.196(18) &  &  & 0.2086(30) & 0.230(39) &  \\
\nuc{72}{Se} & 0.183(20) & 0.190(15) &  &  &  &  \\
\nuc{78}{Kr} & 0.674(33) & 0.659(34) &  & 0.601(30) &  &  \\
\nuc{80}{Kr} & 0.396(26) &  &  & 0.388(20) &  &  \\
\nuc{86}{Sr} & 0.1400(70) &  &  & 0.109(16) &  &  \\
\nuc{88}{Sr} & 0.0909(47) &  &  & 0.104(15) & 0.0882(59) & 0.094(12) \\
\nuc{90}{Zr} & 0.086(13) &  &  &  &  & 0.057(10) \\
\nuc{92}{Mo} & 0.102(18) &  &  & 0.1052(60) &  &  \\
\nuc{94}{Mo} & 0.1959(95) &  &  & 0.2145(98) &  &  \\
\nuc{96}{Mo} & 0.269(15) &  &  & 0.283(14) &  &  \\
\nuc{96}{Ru} & 0.231(10) &  &  & 0.244(12) &  &  \\
\nuc{98}{Ru} &  & 0.413(14) &  & 0.395(19) &  &  \\
\nuc{106}{Pd} &  &  &  & 0.662(37) &  & 0.608(49) \\
\nuc{108}{Pd} &  &  &  & 0.762(50) &  & 0.807(40) \\
\nuc{110}{Pd} &  & 0.838(42) &  & 0.879(60) &  & 0.850(44) \\
\nuc{110}{Cd} &  &  &  & 0.438(21) &  & 0.450(35) \\
\nuc{114}{Cd} &  &  &  & 0.536(25) &  & 0.538(28) \\
\nuc{112}{Sn} & 0.195(14) &  &  & 0.241(11) &  &  \\
\nuc{114}{Sn} & 0.185(12) &  &  & 0.233(12) &  &  \\
\nuc{116}{Sn} &  &  &  & 0.2135(50) & 0.194(17) & 0.199(27) \\
\nuc{118}{Sn} &  &  &  & 0.2052(40) & 0.218(20) & 0.172(14) \\
\nuc{120}{Sn} & 0.1982(98) &  &  & 0.202(10) & 0.186(22) & 0.136(22) \\
\nuc{124}{Sn} & 0.1414(90) &  &  & 0.1651(41) & 0.164(22) &  \\
\nuc{122}{Te} &  &  &  & 0.647(30) & 0.64(10) &  \\
\nuc{124}{Te} &  &  &  & 0.566(28) & 0.616(88) &  \\
\nuc{124}{Xe} &  & 0.978(66) &  & 0.98(11) &  &  \\
\nuc{130}{Ba} &  & 1.111(55) &  & 1.157(79) &  &  \\
\nuc{138}{Ba} &  &  &  & 0.229(11) & 0.241(17) &  \\
\nuc{144}{Ba} &  & 1.02(18) & 1.020(55) &  &  &  \\
\nuc{140}{Ce} & 0.307(16) &  &  & 0.291(15) & 0.308(25) &  \\
\nuc{142}{Ce} &  &  &  & 0.467(23) &  & 0.457(23) \\
\nuc{142}{Nd} & 0.265(13) &  &  & 0.272(19) & 0.254(19) & 0.308(49) \\
\nuc{146}{Nd} &  &  &  & 0.750(38) &  & 0.655(44) \\
\nuc{144}{Sm} & 0.228(74) &  &  & 0.263(13) &  &  \\
\nuc{152}{Sm} &  &  & 3.4613(23) & 3.43(17) &  & 3.41(17) \\
\nuc{154}{Gd} &  &  & 3.874(16) & 3.79(19) &  &  \\
\nuc{156}{Gd} &  &  & 4.77(11) & 4.53(23) &  &  \\
\nuc{158}{Gd} &  &  & 5.08(17) & 5.09(25) &  &  \\
\nuc{160}{Gd} &  &  & 5.21(11) & 5.28(26) &  &  \\
\nuc{156}{Dy} &  &  & 3.66(22) & 3.74(19) &  &  \\
\nuc{158}{Dy} &  &  & 4.65(24) & 4.67(23) &  &  \\
\nuc{162}{Dy} &  &  & 5.27(26) & 5.03(26) &  &  \\
\nuc{162}{Er} &  &  & 5.61(54) & 4.95(25) &  &  \\
\nuc{164}{Er} &  &  & 5.25(21) & 5.32(27) &  &  \\
\nuc{166}{Er} &  &  & 5.78(16) & 5.68(24) &  &  \\
\nuc{168}{Er} &  &  & 5.681(59) & 5.98(23) &  &  \\
\nuc{172}{Yb} &  &  & 6.16(15) & 5.96(29) &  &  \\
\nuc{174}{Yb} &  &  & 5.91(31) & 5.84(29) &  &  \\
\nuc{174}{Hf} &  &  & 5.89(22) & 5.30(35) &  &  \\
\nuc{178}{Hf} &  &  & 4.797(94) & 4.66(23) &  &  \\
\nuc{180}{Hf} &  &  & 4.6471(30) & 4.58(22) &  &  \\
\nuc{182}{W} &  &  & 4.091(61) & 4.09(16) &  &  \\
\nuc{184}{W} &  &  & 3.57(11) & 3.89(19) &  &  \\
\nuc{186}{Os} &  &  & 3.060(72) & 3.11(25) &  &  \\
\nuc{188}{Os} &  &  & 2.449(59) & 2.69(13) &  &  \\
\nuc{190}{Os} &  & 1.98(21) & 3.09(72) & 2.37(11) &  &  \\
\nuc{192}{Os} &  &  &  & 2.04(10) &  & 2.01(10) \\
\nuc{192}{Pt} &  &  & 2.002(95) & 1.931(90) &  &  \\
\nuc{194}{Pt} &  & 1.428(68) &  & 1.683(80) &  &  \\
\nuc{196}{Pt} &  & 1.348(70) &  & 1.418(68) &  & 1.429(71) \\
\nuc{198}{Pt} &  & 1.030(52) &  & 1.098(50) &  &  \\
\nuc{198}{Hg} &  &  & 1.084(83) & 0.9600(68) & 0.74(10) &  \\
\nuc{208}{Pb} &  &  &  &  & 0.263(18) & 0.312(16) \\
\nuc{230}{Th} &  &  & 8.12(21) & 8.22(68) &  &  \\
\nuc{234}{U} &  &  & 9.92(50) & 10.57(55) &  &  \\
\nuc{236}{U} &  &  & 10.78(28) & 11.68(58) &  &  \\
\nuc{240}{Pu} &  &  & 13.12(39) & 13.11(65) &  &  \\
\hline \hline
\end{longtable}
\end{center}

\clearpage
\setcounter{figure}{0}
\renewcommand{\thefigure}{B\arabic{figure}}
\section{Additional $z$-Value Histograms}\label{app:hist}
\begin{figure}[h!]
\centering
\begin{subfigure}{0.45\textwidth}
  \centering
  \includegraphics[width=\linewidth]{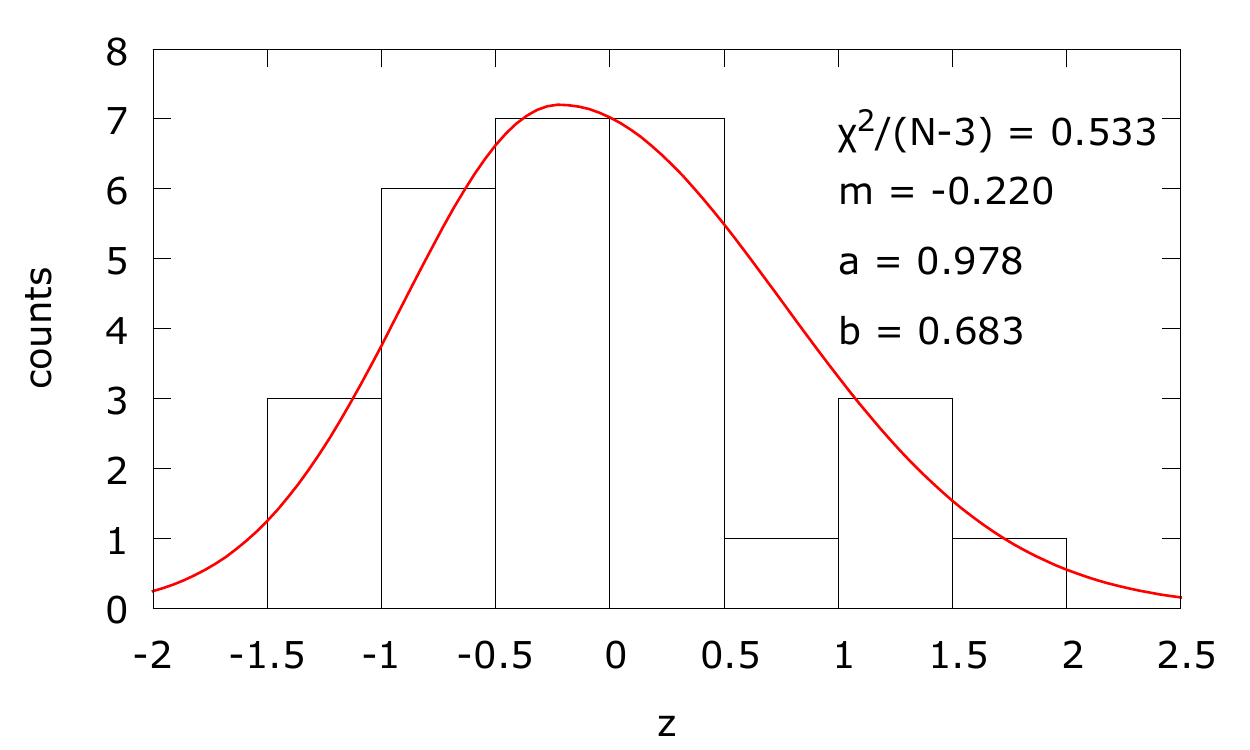}
  \caption{\label{fig:cedc}}
\end{subfigure}
\begin{subfigure}{0.45\textwidth}
  \centering
  \includegraphics[width=\linewidth]{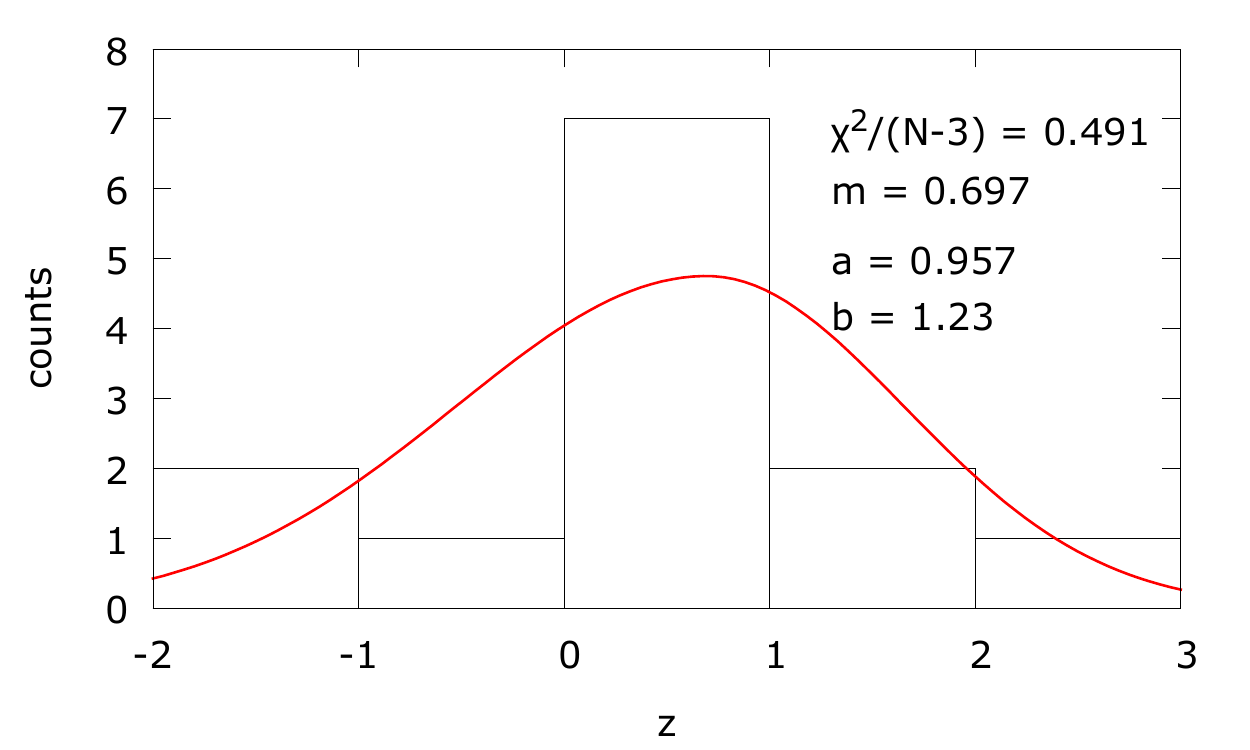}
  \caption{\label{fig:cerdds}}
\end{subfigure}
\begin{subfigure}{0.45\textwidth}
  \centering
  \includegraphics[width=\linewidth]{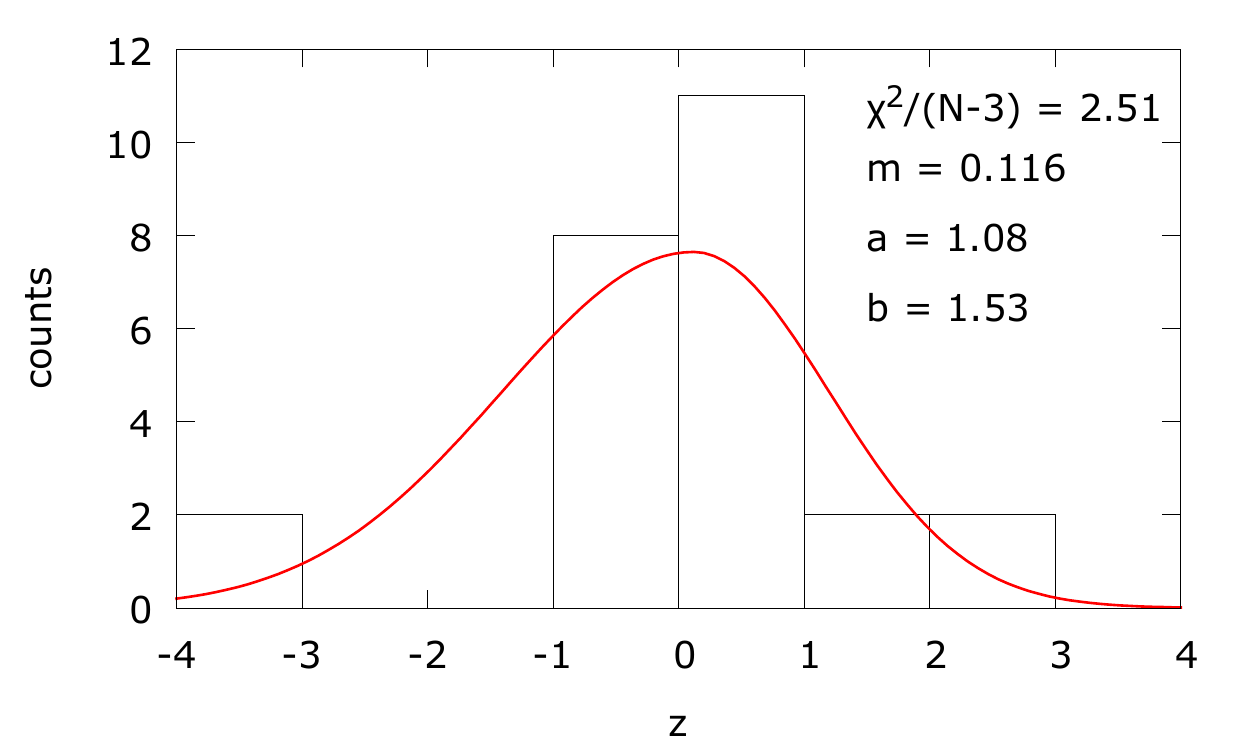}
  \caption{\label{fig:cenrf}}
\end{subfigure}
\caption{(a) Histograms of the $z$ values computed for the CE/DC method pair, together with the asymmetric
normal distribution fit. (b) Same as (a), but for the CE/RDDS pair. (c) Same as (a), but for the
CE/\ggp\ pair.\label{fig:apphists}}
\end{figure}

\clearpage
\setcounter{table}{0}
\renewcommand{\thetable}{C\arabic{table}}
\section{Nuclide Indices}
\begin{center}
\begin{table}
\caption{List of nuclides corresponding to the indices given on the x-axes of the plots in 
Fig.~\ref{fig:diffs}.\label{tab:indicies}}
\begin{tabular}{cccccc|cc}
\hline\hline
Index & Fig.~\ref{fig:diffs_dsa} & Fig.~\ref{fig:diffs_rdds} & Fig.~\ref{fig:diffs_dc} & Fig.~\ref{fig:diffs_gg} & Fig.~\ref{fig:diffs_ee} &Index & Fig.~\ref{fig:diffs_dsa} \\ \hline
1 & \nuc{18}O & \nuc{18}O & \nuc{152}Sm & \nuc{24}Mg & \nuc{18}O & 33 & \nuc{92}Mo \\
2 & \nuc{18}Ne & \nuc{22}Ne & \nuc{154}Gd & \nuc{26}Mg & \nuc{22}Ne & 34 & \nuc{94}Mo \\
3 & \nuc{20}Ne & \nuc{24}Mg & \nuc{156}Gd & \nuc{28}Si & \nuc{24}Mg & 35 & \nuc{96}Mo \\
4 & \nuc{22}Ne & \nuc{46}Ti & \nuc{158}Gd & \nuc{32}S & \nuc{26}Mg & 36 & \nuc{96}Ru \\
5 & \nuc{24}Mg & \nuc{78}Kr & \nuc{160}Gd & \nuc{46}Ti & \nuc{28}Si & 37 & \nuc{112}Sn \\
6 & \nuc{26}Mg & \nuc{98}Ru & \nuc{156}Dy & \nuc{48}Ti & \nuc{32}S & 38 & \nuc{114}Sn \\
7 & \nuc{28}Si & \nuc{110}Pd & \nuc{158}Dy & \nuc{52}Cr & \nuc{44}Ca & 39 & \nuc{120}Sn \\
8 & \nuc{32}S & \nuc{124}Xe & \nuc{162}Dy & \nuc{56}Fe & \nuc{52}Cr & 40 & \nuc{124}Sn \\
9 & \nuc{34}S & \nuc{130}Ba & \nuc{162}Er & \nuc{58}Ni & \nuc{54}Fe & 41 & \nuc{140}Ce \\
10 & \nuc{36}Ar & \nuc{190}Os & \nuc{164}Er & \nuc{60}Ni & \nuc{56}Fe & 42 & \nuc{142}Nd \\
11 & \nuc{40}Ar & \nuc{194}Pt & \nuc{166}Er & \nuc{64}Zn & \nuc{58}Ni & 43 & \nuc{144}Sm \\
12 & \nuc{44}Ca & \nuc{196}Pt & \nuc{168}Er & \nuc{66}Zn & \nuc{60}Ni &  &  \\
13 & \nuc{48}Ti & \nuc{198}Pt & \nuc{172}Yb & \nuc{68}Zn & \nuc{62}Ni &  &  \\
14 & \nuc{50}Ti &  & \nuc{174}Yb & \nuc{72}Ge & \nuc{64}Ni &  &  \\
15 & \nuc{50}Cr &  & \nuc{174}Hf & \nuc{88}Sr & \nuc{64}Zn &  &  \\
16 & \nuc{52}Cr &  & \nuc{178}Hf & \nuc{116}Sn & \nuc{66}Zn &  &  \\
17 & \nuc{54}Fe &  & \nuc{180}Hf & \nuc{118}Sn & \nuc{68}Zn &  &  \\
18 & \nuc{56}Fe &  & \nuc{182}W & \nuc{120}Sn & \nuc{88}Sr &  &  \\
19 & \nuc{58}Ni &  & \nuc{184}W & \nuc{124}Sn & \nuc{106}Pd &  &  \\
20 & \nuc{60}Ni &  & \nuc{186}Os & \nuc{122}Te & \nuc{108}Pd &  &  \\
21 & \nuc{62}Ni &  & \nuc{188}Os & \nuc{124}Te & \nuc{110}Pd &  &  \\
22 & \nuc{64}Ni &  & \nuc{190}Os & \nuc{138}Ba & \nuc{110}Cd &  &  \\
23 & \nuc{64}Zn &  & \nuc{192}Pt & \nuc{140}Ce & \nuc{114}Cd &  &  \\
24 & \nuc{66}Zn &  & \nuc{198}Hg & \nuc{142}Nd & \nuc{116}Sn &  &  \\
25 & \nuc{68}Zn &  & \nuc{230}Th & \nuc{198}Hg & \nuc{118}Sn &  &  \\
26 & \nuc{70}Zn &  & \nuc{234}U &  & \nuc{120}Sn &  &  \\
27 & \nuc{70}Ge &  & \nuc{236}U &  & \nuc{142}Ce &  &  \\
28 & \nuc{72}Ge &  & \nuc{240}Pu &  & \nuc{142}Nd &  &  \\
29 & \nuc{78}Kr &  &  &  & \nuc{146}Nd &  &  \\
30 & \nuc{80}Kr &  &  &  & \nuc{152}Sm &  &  \\
31 & \nuc{86}Sr &  &  &  & \nuc{192}Os &  &  \\
32 & \nuc{88}Sr &  &  &  & \nuc{196}Pt &  & \\ \hline\hline
\end{tabular}
\end{table}
\end{center}

\newpage
{\large \textbf{References}}
\bibliographystyle{model1a-num-names}

\end{document}